\documentclass[twocolumn,10pt,showpacs,superscriptaddress]{revtex4-1}

\usepackage{graphicx}
\usepackage{amssymb,amsfonts,amsmath}
\usepackage{epstopdf}

    \usepackage{multirow}

\begin{document}

\title{Soft Matter Perspective on Protein Crystal Assembly}

\author{Diana Fusco}
\affiliation{Department of Physics, University of California, Berkeley, California 94720, USA}
\affiliation{Department of Integrative Biology, University of California, Berkeley, California 94720, USA}
\affiliation{Program in Computational Biology and Bioinformatics, Duke University, Durham, NC 27708}
\affiliation{Department of Chemistry, Duke University, Durham, NC 27708}

\author{Patrick Charbonneau}
\affiliation{Program in Computational Biology and Bioinformatics, Duke University, Durham, NC 27708}
\affiliation{Department of Chemistry, Duke University, Durham, NC 27708}
\affiliation{Department of Physics, Duke University, Durham, NC 27708}

\begin{abstract} 
Crystallography may be the gold standard of protein structure determination, but obtaining the necessary high-quality crystals is also in some ways akin to prospecting for the precious metal. The tools and models developed in soft matter physics to understand colloidal assembly offer some insights into the problem of crystallizing proteins. This topical review describes the various analogies that have been made between proteins and colloids in that context. We highlight the explanatory power of patchy particle models, but also the challenges of providing guidance for crystallizing specific proteins. We conclude with a presentation of possible future research directions. This article is intended for soft matter scientists interested in protein crystallization as a self-assembly problem, and as an introduction to the pertinent physics literature for protein scientists more generally.
\end{abstract}

\maketitle


\section{Introduction}
Biological macromolecules are central to life processes. Although some of these processes can be characterized at a relatively coarse scale, more often than not a microscopic understanding of the structure and dynamics of the involved biomolecules is also essential\footnote{This phenomenon is sometimes playfully referred to as the revenge of structural biology.}. Proteins, for instance, interact with each other and their environment through fine-tuned structural features so as to perform their biological functions~\cite{blundell:2002,kuhn:2002,blundell:2004,tickle:2004,kitchen:2004}; reciprocally, protein malfunction is often due to structural defects, and may result in diseases~\cite{congreve:2005}. Proteins also represent some of the most sophisticated nano-machines known, having been shaped by natural selection. From an engineering perspective, few other systems offer more insights into designing devices on that size scale~\cite{huebsch:2009}. Our limited knowledge of protein structures thus limits our comprehension of biological phenomena, our ability to discover new drugs, and our design of bio-inspired materials~\cite{mcpherson:1999,chayen:2008,Chayen2009}. For that reason, sizable research efforts are being expended on extracting detailed protein structures and dynamics~\cite{Morange:2006,Khafizov:2014}.

As a field, structural biology mostly relies on protein structures obtained from diffraction-based methods. Since X-ray crystallography enabled the structure determination of myoglobin and hemoglobin in the 50's, both the amount and quality of structural information have greatly increased~\cite{jones:2014}. Protein crystallography, however, requires protein crystals, whose obtention is often the limiting experimental step~\cite{chen:2004,terwilliger:2009}. Expertise from many different fields -- from computational biology and robotics to surfactant science -- has thus been brought to bear on the problem~\cite{mcpherson:1999}. 
This article specifically reviews the contribution of soft matter physics to understanding the crystallization of globular proteins\footnote{The assembly of membrane proteins typically relies on different physical processes~\cite{caffrey:2009}. These processes are generally beyond the scope of this review, although we get back to this issue in the conclusion.}. Although the scientific conversation between the soft matter and structural biology communities has thus far mostly focused on providing a physical rationale for experimental observations, recent conceptual advances and an increased back and forth between theory and experiment suggest that richer exchanges may soon become the norm. In the following, we recapitulate how this development has come to be. We present in Sect.~\ref{sec:diffration} and~\ref{sec:solubility} an overview of the basics of protein crystallography and protein crystal assembly, respectively, in order to motivate the practical importance of physical insights. We then discuss various isotropic (Sect.~\ref{sec:isotropic}) and patchy (Sect.~\ref{sec:patchy}) models of protein assembly. Section~\ref{sec:conclusion} concludes with a discussion of possible future research directions. 

Note that we aim here to provide a reasonably broad physical description of protein crystallization and its constraints. Our target readership are soft matter scientists with either a novel or a renewed interest in the problem. We thus describe some more elementary aspects of the underlying structural biology and crystallization technology. We intend, however, the later sections to also serve as an introduction to the soft matter literature for protein scientists more generally.

\section{Diffraction and protein crystallization}
\label{sec:diffration}
Proteins are encoded in DNA as a series of base pairs, which are then translated into a sequence of amino acids, forming the primary structure of the molecules. To perform its functions, a protein must also hierarchically fold into its secondary and tertiary structures. It is the properties of the resulting three-dimensional object that largely determine how the molecule interacts with its environment. This connection between function and shape is at the root of both structural biology and structural genomics.

Advances in high-throughput sequencing have markedly increased the number of protein-encoding genes and thus of known primary structures. Knowledge of the amino acid sequence alone, however, currently provides but limited insights into the higher-order structure of a protein. Predicting a protein's full tertiary structure from its sequence is indeed a remarkably complex task~\cite{Dill:2012}. Because for most purposes protein structures cannot be inferred or calculated, they must be determined experimentally. The classical and most frequently used approaches for doing so are X-ray and neutron crystallography. These solid-phase diffraction methodologies can handle proteins containing several thousands of amino acids and reach sub-atomic resolution ($<1.5$ \AA). By contrast, in spite of important methodologically advances, nuclear magnetic resonance (NMR) still cannot resolve the structure of solvated proteins containing more than two to three hundred amino acids, i.e., a few tens of kilodaltons~\cite{Grzesiek2009,kerfah:2015}. NMR-resolved structures thus represent only a small (although increasing) fraction of the protein data bank (PDB). Recent advances have also pushed single-particle electron cryomicroscopy near atomic resolution~\cite{doerr:2014,li:2013,bai:2013}. However, (sub-)atomic resolution may remain physically unattainable because of radiation damage, beam-induced movement and charging of the sample~\cite{bai:2013}. The classical diffraction methodologies are thus likely to remain the reference for the foreseeable future. Interestingly, new diffraction-based techniques are also under development. Electrons, which scatter fairly strongly from molecules, can provide a diffracted image of a protein's Coulomb potential~\cite{Yonekura:2015}, and the intense X-ray pulses from free-electron lasers (FEL) provide high-resolution structural information from relatively small crystals~\cite{barends:2014}.

Whatever the chosen diffraction approach may be, three main steps must be performed in order to prepare a sample: (i) the protein must be expressed in a sufficient amount, which is typically achieved by using plasmids in cultured bacterial cells; (ii) the protein must be purified, in order to isolate it from the biological medium used to express it; and (iii) the purified protein molecules must be assembled in an ordered and well-packed crystallite, whose minimal size depends on the scattering intensity of the diffracted radiation (from tens of nanometers or less on the side for FEL to millimeters for neutron beams~\cite{stevenson:2014,blakeley:2008}). Although each of these steps presents several experimental challenges, crystallization is by far the most troublesome~\cite{mcpherson:1999,chen:2004,chayen:2008,Chayen2009}. Under standard biological conditions, most proteins do not easily crystallize. They likely evolved to limit aggregation that interferes with their normal biological function~\cite{janin:1995,doye:2004} (with a few notable exceptions~\cite{mcpherson:1999,doye:2006}). Some neurodegenerative disorders, such as Alzheimer's and Parkinson's diseases, have indeed been linked to protein solubilization failure resulting in unwanted aggregation~\cite{ross:2004}, and cataract formation can directly involve the crystallization of eye-lens proteins~\cite{kmoch:2000,pande:2001,evans:2004,li:2008}. Hence, in order to promote their periodic assembly, proteins must be placed in chemical conditions that are typically far from those encountered in biological systems, yet must not result in a loss of secondary or ternary structure.

Even within those strict constraints, the chemical space within which to locate conditions that promote crystallization is remarkably vast. Exhaustive searches are simply beyond experimental reach~\cite{newman:2012}. Crystallographers have thus developed chemical screens summarizing conditions that have been previously successful. For globular proteins, these are aqueous solutions containing various combinations of three families of co-solutes: inorganic salts, polymers (e.g., polyethylene glycol), and small organic molecules~\cite{mcpherson:1999}. Technological advances in automatizing the experimental process currently allow for up to thousands of these crystallization cocktails to be tested at once~\cite{Bruno:2014}, but even that number is only a minute fraction of the full spectrum of chemical possibilities. 

More crucially, existing screens are far from guaranteeing crystal formation. Despite accrued experience and improved experimental techniques, successful crystallization indeed remains the exception rather than the rule. On average, only 0.04\% of crystallization experiments generate good-quality crystals, which makes them quite time consuming and expensive to obtain~\cite{chen:2004}. Membrane proteins are even harder to crystallize~\cite{caffrey:2003}.  Yet as long as diffraction-based methods remain the most desirable way to determine protein structures, a better understanding of the physical mechanisms underlying protein crystal assembly is the only possible path towards a high-throughput scheme comparable to next-generation DNA sequencing~\cite{Morange:2006,Stevens:2013}. Brute-force approaches are unlikely to succeed on their own.

\section{Solubility phase diagrams and protein crystallization}
\label{sec:solubility}
\begin{figure}
\centering
\includegraphics[width=0.4\textwidth]{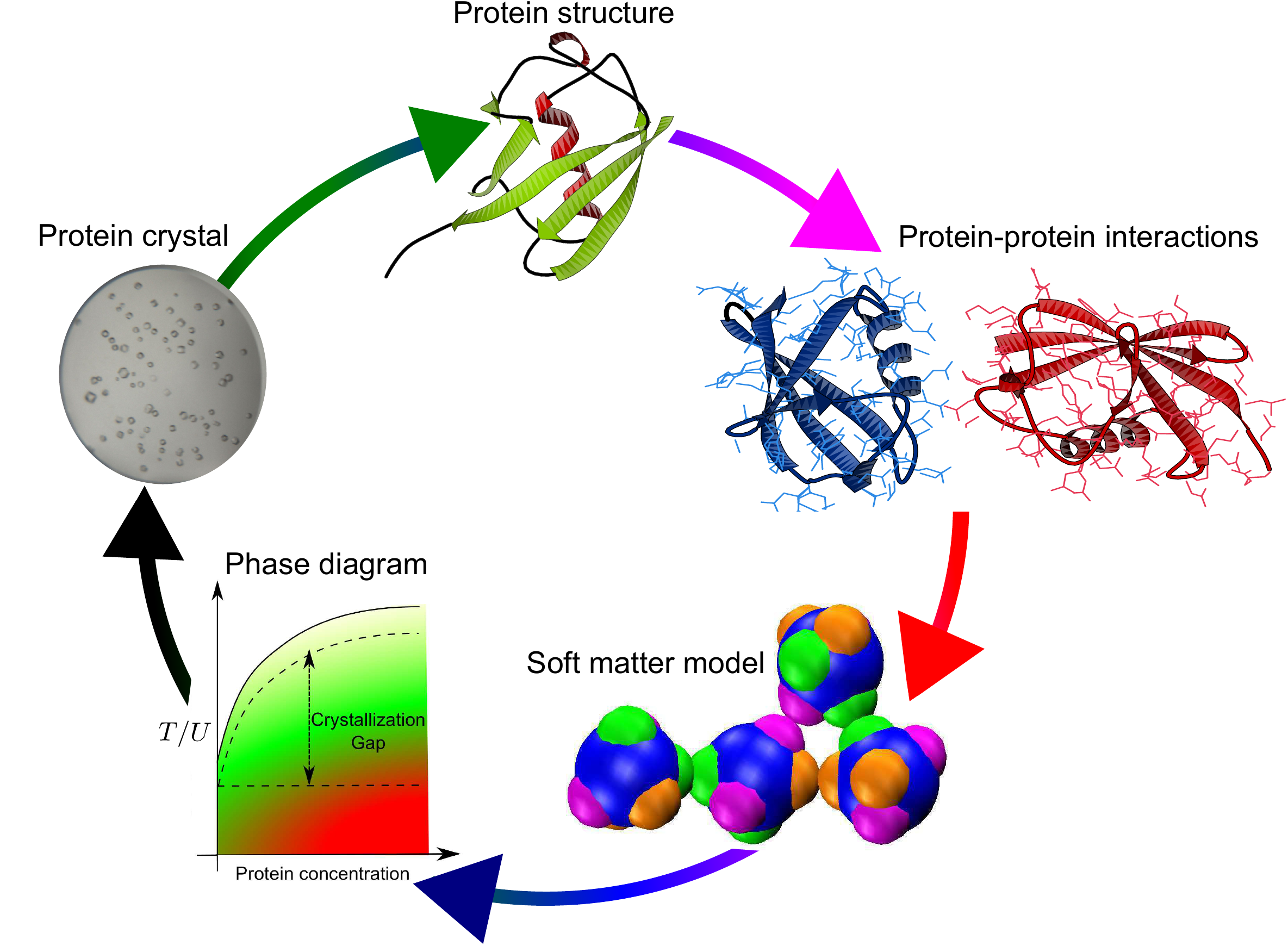}
\caption{Schematic description of the vicious circle of protein crystallization. Crystals are needed to obtain protein structures through diffraction (green arrow), but these crystals would be more easily obtained if experimental phase diagrams were known (black arrow). These phase diagrams are, however, expensive to determine. Phase diagrams of coarse-grained protein models may be accessible via simulations (blue arrow), but an appropriate protein model should stem from a detailed understanding of protein-protein interactions (red arrow), the computation of which requires some knowledge of the protein structure (purple arrow). By contextualizing the observed protein behavior, however, soft matter models could provide guidelines to refine and iterate the process.}
\label{fig:circle}
\end{figure}

From a thermodynamic viewpoint, protein crystallization is a standard phase transition whereat the chemical potential of protein molecules dispersed in an aqueous solvent is equal to that of those ordered in a crystal. The transition is in some ways similar to water freezing into ice, suggesting that liquid-state descriptions may provide useful microscopic insights. Understanding protein crystallization, or any other phase transition, should indeed follow from knowing the (effective) interactions between protein molecules~\cite{mcpherson:1999,derewenda:2010,anderson:2006,saridakis:2009}. In other words, a detailed description of protein-protein interactions should also provide the protein phase behavior (Fig.~\ref{fig:circle})~\cite{dumetz:2009}. Yet, although the underlying forces through which proteins interact, i.e., hydrogen bonding, van der Waals, hydrophobic, and electrostatic interactions, are individually fairly well characterized~\cite{israelachvili:1991,Chandler2005,gunton:2007,granick:2008,Berne2009}, their collective contribution to protein assembly is much less well understood~\cite{vekilov:2002,devedjiev:2015}. A way to somehow coarsen microscopic details into a simpler, effective description of protein pair interactions is thus needed.

Phase diagrams capture graphically the regions of parameter space over which different phases of matter are thermodynamically stable. For mixtures, the phase diagram is multi-dimensional, but in some systems two-dimensional projections on the temperature-concentration plane of a key component, e.g., the protein, recover a significant fraction of that information. For crystallization cocktails, however, this projection typically involves rescaling the temperature axis by an aggregate function of the solution conditions. Co-solutes can indeed change the energy scale over which protein molecules interact with one another.
For many proteins, this projection displays phases that are analogous to those of a single-component system, for which the two-dimensional projection is complete. Crystal, liquid, vapor, and supercritical fluid regimes are indeed identifiable. The vapor phase corresponds to a low-concentration of proteins in solution and the liquid to a high-concentration. The two phases are separated by a first-order transition that terminates at a second-order critical point, above which a supercritical fluid is observed. Because protein solutions are mixtures of proteins, water, and other additives, however, the typical properties of phases in a protein system may significantly differ from what is typically observed in simple fluids. A protein crystal, for instance, contains on average more than 40\% water~\cite{Matthews:1968} and may embed co-solutes in fractions that differ from what is left in the crystallization cocktail.
 
Phase diagrams for about a dozen proteins, including lysozyme, $\gamma$-crystallin, insulin, and myoglobin, have been experimentally determined under different solution conditions~\cite{Malfois:1996,lomakin:1996,muschol:1997,stradner:2005,mcmanus:2007,talreja:2010,Gogelein2012}. Intriguingly, these phase diagrams have in common a topology that is qualitatively different from that of simple liquids (Fig.~\ref{fig:PD})~\cite{hansen:2006}. In simple liquids, the first-order phase transition between a vapor and a liquid phase terminates at a stable second-order critical point that lies \emph{above} the crystal solubility curve, whereas in proteins the critical point is typically situated \emph{below} the solubility line. The protein gas-liquid binodal and critical point are, therefore, metastable with respect to gas-crystal coexistence (Fig.~\ref{fig:PD}). 

It is also empirically found that successful crystallization typically is most common when a protein solution drop is prepared in the supersaturated region between the solubility line and the metastable critical point -- a region that is sometimes called the nucleation zone or the crystallization gap (Fig.~\ref{fig:PD}A)~\cite{mcpherson:1999,asherie:2004,asherie:2012,sleutel:2015}. If it is prepared at conditions below the critical point, the system has a propensity to aggregate in a disordered, percolating network, i.e., a gel; the gelation probability increases with quench depth~\cite{thomson:1987,dumetz:2008}. Such aggregation, although thermodynamically metastable with respect to crystal assembly, forms much more rapidly and is often long-lived. It thus reduces the likelihood of successful protein crystallization. 

Based on this description, given a protein phase diagram, protein crystallization should be easily achieved (Fig~\ref{fig:circle}). Unfortunately, experimental determination of protein phase diagrams (considering the vast number of possible co-solutes) is a lot more time- and resource-consuming than even the most ambitious crystallization screen. Hence, although physically elegant, a direct application of thermodynamics principles is of limited practical relevance. This picture is also overly reductive because a protein's structure may also respond to changing solution conditions. In order to gain useful insights about protein crystal assembly, one must therefore extract the key features of their phase diagrams from a limited amount of information. Because atomistic models of crystal assembly would be computationally intractable (simulating a simple protein can be quite demanding, let alone hundreds of them), coarse-grained models are the only viable option to fill in the missing information (Fig.~\ref{fig:circle}). Conveniently, self-assembly has been heavily studied in related models over the last couple of decades.

\section{Isotropic models of protein phase behavior}
\label{sec:isotropic}
The field of soft matter physics has grown from a desire to understand the physics behind the assembly of squishy, mesoscale objects, such as polymers, liquid crystals, grains, and cells. From the self-assembly of these objects  emerges a complex yet often universal array of material behaviors, including glass formation and jamming, ordered microphases, and tissue growth. Soft materials are also particularly appealing because a large part of their complexity can be captured by purely classical models, and because the relevant range of these model's parameters is much wider than for atomic-scale simple liquids.

The study of the rich and robust phenomenology of colloidal suspensions, which in some ways epitomizes the field of soft matter, opens the door to understanding protein assembly. In the mid-80's, suspensions of purely repulsive colloidal particles were first observed to crystallize similarly to classical hard spheres~\cite{Pusey:1986}. Adding a depletant, i.e., a soluble and inert co-solute much smaller than the colloids, was understood to result in a net pair attraction between particles~\cite{Asakura:1954,Vrij:1976}, and, based on the law of corresponding states~\cite{hansen:2006}, was expected to result in the presence of a gas-liquid coexistence region. Experiments showed instead that adding a depletant unavoidably results in gel formation~\cite{gast:1983}. 

\begin{figure*}[htb]
\centering
\includegraphics[width=0.9\textwidth]{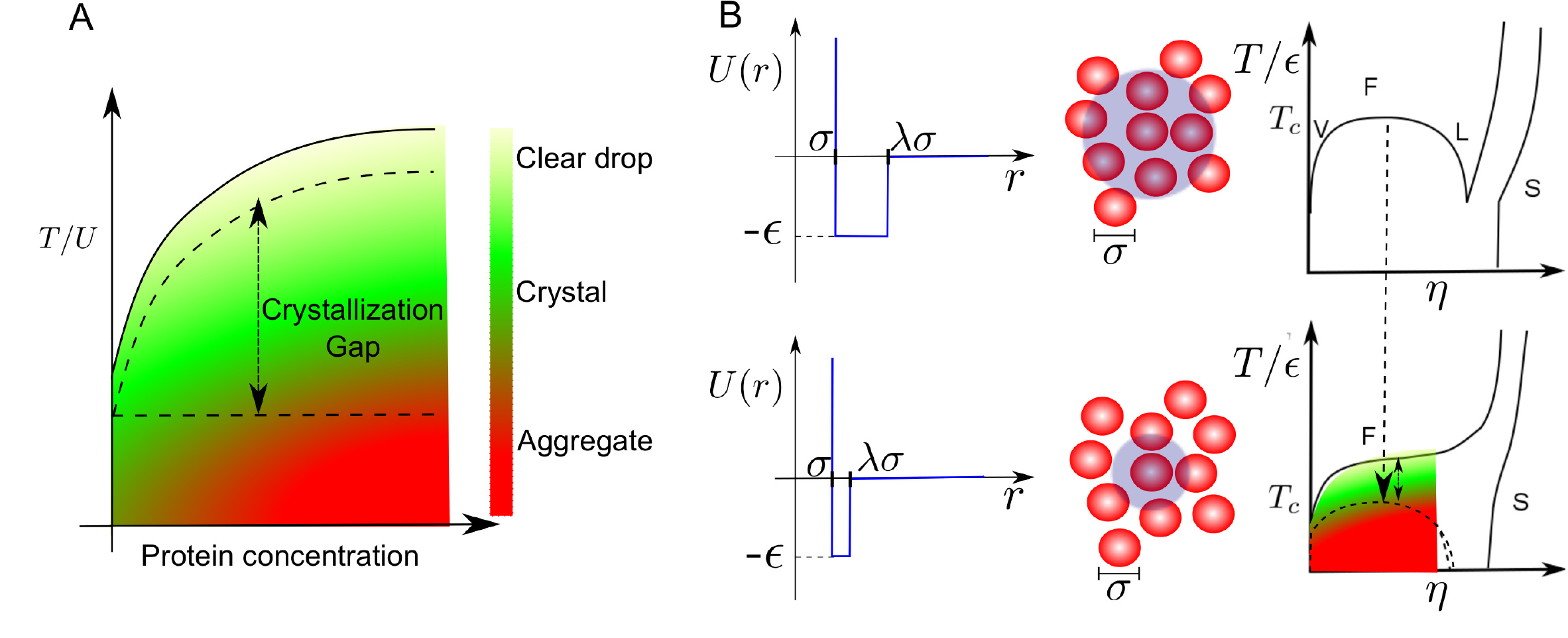}
\caption{A: Typical topology of a temperature $T$-protein concentration phase diagram, where the effective interaction $U$ can be tuned by changing solution conditions, e.g. changing salt concentration and pH, adding depletants, etc. At low saturation (yellow in the color bar), the drive to crystallize is insufficient to generate crystals within reasonable experimental time. At high saturation (red in the color bar), the drive to assemble is so strong that molecules form amorphous aggregates. In between lies the nucleation zone (or crystallization gap), where crystal assembly is possible (green in the color bar). By tuning $T$ or $U$ experimentally, the vertical axis can be rescaled and the saturation level controlled. B: The radial range of attraction, $\lambda\sigma$, of the pair interaction potential, $U(r)$, affects the topology of the scaled temperature-packing fraction, $T/\epsilon$--$\eta$, phase diagram. The situation is here illustrated for square-well fluids, but is fairly independent of the precise form of $U(r)$. In simple liquids, the relative interaction range is sufficiently large, $\lambda\gtrsim 1.25$, to make the liquid state entropically stable (top half). In colloids and proteins, the short-range attraction, i.e., for $\lambda\lesssim 1.25$, the reduced liquid entropy lowers the critical point below the solubility line (bottom half).}
\label{fig:PD}
\end{figure*}

A possible resolution to this discrepancy emerged from the work of Lekkerkerker and Frenkel~\cite{Lekkerkerker:1992,Hagen1994}, who found that decreasing the attraction range depresses the gas-liquid coexistence region more than it lowers the crystal solubility curve (Fig.~\ref{fig:PD}B). For particles with a square-well attraction range $\lambda\sigma\lesssim 1.25\sigma$, relative to particles of diameter $\sigma$, the critical point thus falls below the solubility curve. Different models with short-range attraction confirmed the qualitative robustness of this result~\cite{Asherie:1996,Tavares:1997,Miller:2003,Liu:2005,Pagan:2005,lopezrendon:2006,Largo:2008,fortini:2008}, and its universality was synthesized in an extended law of corresponding states~\cite{Noro:2000,Foffi:2006,platten:2015}. 

The change in behavior in going from long- to short-range attraction can be intuitively explained by considering the energy-entropy balance in the liquid and crystal phases. For long-range attraction, there exists a relatively broad concentration-temperature range within which liquid particles are close enough to benefit from each other's attraction while maintaining the high entropy characteristic of disordered configurations (Fig.~\ref{fig:PD}B). The liquid free-energy can therefore be lower than that of the crystal, despite the crystal having a lower potential energy. When the attraction range is reduced, however, particles in the liquid have to be much closer to each other in order to feel their attraction (Fig.~\ref{fig:PD}B). This constraint drastically reduces the number of low-energy configurations, and thus the liquid entropy. By contrast, for interaction ranges as low as $\lambda\sigma=1.10\sigma$, which is comparable to the interparticle distance in a hard-sphere crystal near melting~\cite{hansen:2006}, the crystal entropy remains almost unchanged. As a result, the liquid free-energy raises above that of the crystal and the liquid phase ceases to be thermodynamically stable (Fig.~\ref{fig:PD}B).

The gas-liquid coexistence, however, remains metastable; theoretical and experimental work has showed that its dynamical influence does not disappear. Homogeneous systems that are supersaturated near the metastable critical point indeed experience critical fluctuations that lower the barrier to crystal nucleation~\cite{Wolde1997}. In addition, if quenched below the critical point, these systems first undergo a spontaneous phase separation that is well described by spinodal decomposition. It is this process that gets arrested by the sluggish particle dynamics in the dense phase, and results in the emergence of a percolating gel network~\cite{Foffi:2005,Manley:2005,Sastry:2006,Charbonneau:2007,Buzzaccaro:2007,Lu2008}. 

This phase behavior is stunningly reminiscent of that of many proteins, as described in Sect.~\ref{sec:solubility}. The analogy between short-range attractive colloids and proteins, which caught the attention of a number of soft matter scientists in the mid-90's~\cite{lomakin:1996,rosenbaum:1996,Wolde1997,chayen:2005}, is further supported by microscopic evidence. First, many globular proteins are roughly spherical objects with a fairly short-range attraction. Their diameter is of at least few nanometers, while the mechanisms that result in pair attraction once electrostatic repulsion is screened, e.g., hydrogen bonding, salt bridges, and hydrophobicity, extend only over a few angstroms. These attractive forces are thus felt within about $10\%$ of the protein diameter, which falls well within the specification for short-range attraction. Second, George and Wilson observed that the optimal solution conditions for protein crystallization were consistent with being in proximity of a metastable critical point~\cite{george:1994,haas:1998,wilson:2014}. Third, contemporary studies of the PDB did not identify any statistical signatures of a preferred relative orientation of proteins chains in crystals, which was consistent with protein-protein attraction being roughly orientation-independent, i.e., isotropic~\cite{janin:1995b,carugo:1997}. This last suggestion, in particular, would prove to be overly simplistic, as we discuss in Sec.~\ref{sec:patchy}. A productive effort at characterizing short-range isotropic models and subsequently rationalizing the phase diagram~\cite{Pellicane:2004,Pagan:2005} and solution behavior~\cite{stradner:2007,dorsaz:2009,foffi:2014} of specific proteins as well as making generic arguments about heterogeneous nucleation~\cite{chayen:2006,Page:2009} from that perspective nonetheless ensued.

\section{Patchy models of phase behavior}
\label{sec:patchy}
\begin{figure*}[htb]
\centering
\includegraphics[width=0.9\textwidth]{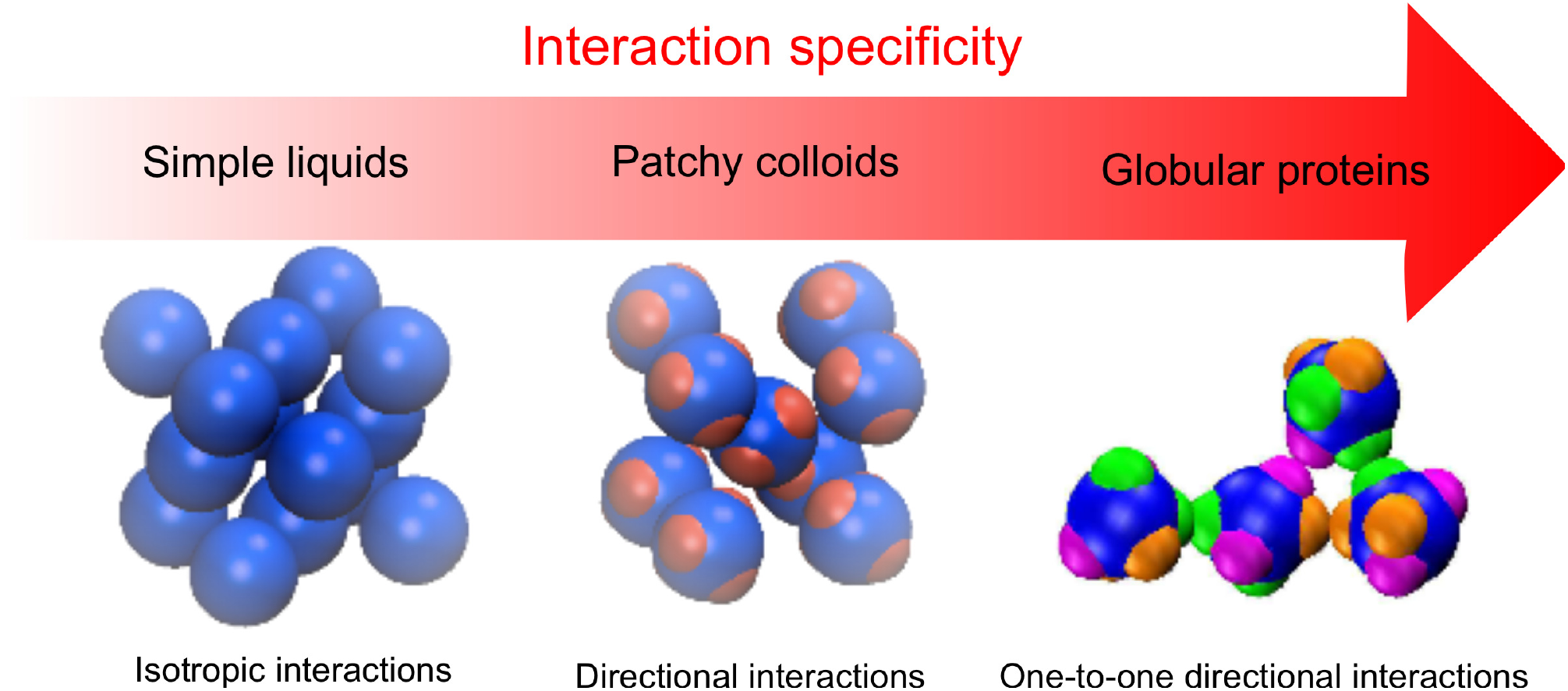}
\caption{Increasing the complexity of schematic models is necessary to accommodate the high interaction specificity that controls protein assembly. Isotropic interactions, which are well-suited to describe simple liquids or depletion interaction in colloids, exclusively depend on the inter-particle distance and assemble in close-packed crystals (left). Thermodynamic stability of low-density crystal requires directionality, which is introduced in patchy models by requiring the alignment of the surface patches for two particles to interact (middle, red patches have to face one another to interact). However, the highly heterogeneous protein surface requires that patches only interact with a single partner patch, and not promiscuously (right, only patches of the same color interact).}
\label{fig:model}
\end{figure*}

Though appealing in their simplicity, colloidal descriptions based on isotropic interactions between proteins miss some of the key phenomenology of globular proteins~\cite{haas:1999,curtis:2001,lomakin:1996,lomakin:1999,mcmanus:2007}. First, isotropic interactions systematically result in densely-packed crystals, such as the face-centered-cubic (FCC) or body-centered-cubic (BCC) lattices (Fig.~\ref{fig:model}), while crystals of biomolecules are fairly \textit{empty}. Their packing fraction $\eta$, i.e., the fraction of space physically occupied by proteins, is between 0.3 and 0.5 (compared to 0.74 and 0.68 for FCC and BCC, respectively), even though the protein chains themselves are essentially close packed~\cite{richards:1974}. Second, the width of the region identified by the metastable vapor-liquid line in experimental protein phase diagrams does not match the expectations for particles with  isotropic interactions~\cite{lomakin:1999,liu:2007,Bianchi2011}. Third, point mutations that change a single amino acid residue at the protein surface can dramatically alter the topology of a protein phase diagram~\cite{mcmanus:2007,mcmanus:2015}, while if attraction were truly isotropic such a small perturbation would be expected to have but a fairly limited impact. 

From a (bio)chemical viewpoint, these observations are far from surprising. Protein surfaces are heterogeneous and complex. There is thus no reason to believe that different relative protein orientations should result in similar interactions. Unlike depletion attractions, hydrogen bonding and salt bridges are strongly directional and localized; they do not evenly cover the whole protein surface. Even the less chemically specific interactions based on hydrophobicity rely on the existence a particularly good structural match (recognition) between two parts of a protein surface, and thus exhibits a strong orientation dependence (see Appendix for a more detailed discussion of specificity). We also now better appreciate, thanks to more careful statistical analysis of the PDB, that crystal contacts are quite different from randomly chosen elements of protein surfaces. They are enriched in glycine and small hydrophobic residues, and depleted in large polar residues with high side-chain entropy, such as lysine and glutamic acids~\cite{Derewenda2009}. A large fraction of glycine and alanine on a protein surface further correlates with a higher probability of the molecule having been crystallized~\cite{price:2009}.

\subsection{Simple Patches}
This evidence drove the realization that directional aspects of protein pair interactions ought to be taken qualitatively into account, in order to obtain reasonable minimal models of proteins (Fig.~\ref{fig:model}). Bonding directionality makes the solution behavior of proteins more akin to that of associative liquids, such as water, than to that of simple (isotropic) liquids, which is a fundamental distinction in liquid-state theory~\cite{hansen:2006}. Nearly coincident with this realization, came the promise of synthetic colloids with an increasing degree of sophistication, such as faceted and DNA-coated particles, which led to a boon of interest in models with directional interactions, as was recently reviewed by Bianchi \textit{et al.}~\cite{Bianchi2011}. Our aim here is not to broadly go over this advance, but to highlight the aspects most important to understanding protein crystallization.

A broad array of schematic models with short-range anisotropic, directional attraction, i.e., \textit{patchy} particle models, have since been studied (Fig.~\ref{fig:model}).
Although the details of patch geometry and interaction parameterization may differ, these patchy models generally have phase diagrams~\cite{Sear1999,lomakin:1999,chang:2004,Gogelein2008,Bianchi2011}, fluid properties~\cite{neal:1998}, and assembly pathways~\cite{hongjun:2009,Whitelam2010,Haxton2012} that are significantly different from those of short-range isotropic models. Most saliently for protein assembly, patchiness further lowers the metastable critical point, stabilizes open geometries such as diamond and cubic crystals, and qualitatively changes the shape of the gas-liquid coexistence region (Table~\ref{table:iso_vs_aniso})~\cite{bianchi:2006,Bianchi2011}. This last point is interesting because experimental phase diagrams of proteins, such as lysozyme and $\gamma$-crystallin, have a binodal whose width and critical packing fraction are much smaller than those of isotropic models~\cite{Gogelein2012,lomakin:1996,lomakin:1999}.
Although patchiness requires the specification of more model parameters and results in a certain loss of universality~\cite{foffi:2007,Bianchi2011}, it also weakens high-order correlations in the fluid structure, which enables the use  of relatively simple liquid-state descriptions, such as Wertheim's theory, for their analysis~\cite{wertheim:1984a,wertheim:1984b,hansen:2006,Bianchi:2007,Bianchi:2008,Bianchi:2011,fusco:2013b}.

Because most of the theoretical studies of patchy particles were motivated by the promises of colloidal synthesis~\cite{glotzer:2007,wang:2012}, they often focused on particles with a couple of patches~\cite{bianchi:2006,charbonneau:2007b,Romano2010,romano:2011}, or on varying the surface coverage of a single patch, i.e., Janus particles~\cite{fantoni:2007,granick:2009,munao:2013}. For particles with one to three patches, small perturbations to the number of patches indeed dramatically affects the liquid behavior, notably enabling the stabilization of empty liquids -- liquid states with a vanishing density~\cite{bianchi:2006,ruzicka:2010}. If this regime is of interest for protein crystal assembly, it is as one to be avoided. Monomeric protein crystals display at least six patches, which is a minimal for mechanical stability~\cite{carugo:1997,krissinel:2010}. The lack of evolutionary pressure for proteins to crystallize indeed results in their crystals having most commonly the lowest-symmetry chiral (protein chains are chiral) point group compatible with that number of contacts, i.e., P$2_12_12_1$~\cite{wukovitz:1995}. Crystals of oligomeric proteins, for which evolution has shaped at least the oligomeric contacts, often assemble in complex asymmetric unit cells with more than ten distinct patches~\cite{dasgupta:1997,minor:2008}. The competition between these interactions can favor the formation of small metastable crystallites, hindering the assembly of defectless crystals.  Patchy models with a larger number of patches have, however, received limited attention thus far~\cite{fusco:2014b}. 

\subsection{Specific Patches}
In addition to being directional, protein-protein interactions are typically specific; there is a one-to-one match between pairs of crystal contacts (Fig.~\ref{fig:model}). Because pairwise attraction between proteins depends sensitively on molecular details, a given patch often interacts exclusively with a single other patch, to the exclusion of all others. 

Interaction specificity has a sizable effect on protein assembly. It obviously affects the liquid entropy and thus the position of the metastable critical point relative to the solubility curve~\cite{hloucha:2001,dixit:2002,dorsaz:2012,fusco:2013a,fusco:2013b}, but more importantly it results in interaction heterogeneity. Because each pair of interacting patches relies on different physicochemical mechanisms, their bonding strength varies~\cite{fusco:2013a}. 
Our recent study of bond energy asymmetry suggest that this factor may play a key role in protein crystallizability. Energy asymmetry alone can indeed result in gel formation due to percolation as in empty liquids, the closing of the crystallization gap, and the restabilization of the critical point above the crystal solubility line~\cite{fusco:2013a}. Future studies will surely expand this list. Note that although no colloidal particle with specific patches as complex as those of proteins have yet been synthesized, patchy coatings of complementary selective DNA strands offer a synthetic gateway for obtaining such surface features~\cite{wang:2012}.

 \begin{table*}
  \caption{Summary of the strengths and weaknesses of short-range attractive models when applied to protein crystallization}
   \label{table:iso_vs_aniso}
   \small{
 \centering
 \begin{tabular}{ccc}
 \hline
 \bf{Model}&\bf{Strengths}&\bf{Weaknesses} \\
 \hline
\multirow{3}{*}{Isotropic}&Minimal number of parameters&Incorrect crystal symmetry\\
&Metastable gas-liquid binodal& Incorrect binodal shape\\
& &Robust to point mutations\\
\hline
 \multirow{3}{*}{Patchy}&Few parameters&Unrealistic pair interactions\\
 &Open crystals&Fixed geometry\\
 & Reasonable bimodal shape& \\
 \hline
\multirow{3}{*}{Specific patches}&Correct crystal symmetry&Many parameters\\
&Realistic pair interactions&Fixed geometry\\
&Tunable for specific proteins&\\
 \hline
 \end{tabular}
 }
 \end{table*}

\subsection{Measuring Patches}
\label{sec:measure}
The loss of generality that accompanies patchiness and specificity leads even schematic models of proteins to require a broad range of system-specific parameters. In order to develop a relevant model for a given protein, a better description of patch energetics and coverage is thus necessary. Although the PDB provides a wealth of structural information about crystal contacts~\cite{Derewenda2009,Price:2011}, relating this information to an effective free energy of interaction in solution is not straightforward. Thus far, the energetics of very few crystal contacts has been characterized. 

An early effort used a phenomenological model to estimate pair interactions from a PDB structure~\cite{hloucha:2001}, in order to reconstruct the phase behavior of bovine chymotrypsinogen. Existing chemical databases do not suffice, however, to generalize this approach with much accuracy. More recent attempts have used all-atom molecular dynamics simulations of protein pairs in solutions in order to extract the angularly-resolved potential of mean force~\cite{Pellicane2008,fusco:2013b,Taudt:2015}. Within the quality of the selected molecular force fields, these simulations offer a reasonable characterization of known crystal contacts. Some of these studies were even able to capture the crystal assembly behavior of various proteins. As long as sufficient structural information about patchiness is available, crystal patch energetics can thus be reconstructed reasonably well (Fig.~\ref{fig:circle}). 

For most systems, however, this information is not known a priori. It is thus problematic to measure the patch characteristics of proteins that have not yet been crystallized or that display patchy interactions that are incompatible with the crystal structure. Because for the vast majority of proteins no structural information is available at all, the former is a genuine difficulty. The latter is also important, because evidence suggests that non-crystallographic contacts may play a kinetic role in protein crystal assembly~\cite{Schmit:2012} and may even render metastable crystal phases kinetically accessible~\cite{fusco:2014b}. 

Yet even if the structure of a protein of interest were fully known, brute force molecular dynamics sampling of the relative surface of a pair of proteins is well beyond computational reach. A hierarchy of methods for identifying candidate attractive patches without complete exhaustion is a more promising way forward. Approaches that consider protein-protein ``docking predictions'' as well as structural homology, followed by testing of these suggestions by higher-precision methods have been used for that purpose~\cite{fusco:2013b}, but it is still early days for the vicious circle of protein crystallization to be broken.

\section{Conclusions and Open questions}
\label{sec:conclusion}
Soft matter has thus far provided a qualitative physical perspective to the problem of protein crystal assembly. In order to push the analogy forward, both qualitative and quantitative improvements are needed.

First, as argued in Sec.~\ref{sec:measure}, more and higher quality information about the protein interactions that give rise to patchiness must be obtained. For guiding the crystallization of protein homologues or for optimizing solution conditions so as to improve crystal quality, this information could be particularly useful. From a computational viewpoint, insights into angularly-resolved potentials of mean force could also be gleaned from large-scale simulations. If a sufficiently large database of interactions were available, it might even be conceivable to use statistical methods to parametrize the patchiness of protein-protein interactions without first extensively simulating the system. From an experimental viewpoint, careful studies of weak protein-protein interactions, as has recently been done for ubiquitin in solution would also be of much help~\cite{Liu:2012}. Studying the structure of transient complexes in solution~\cite{acuner:2011} may not only help identify potential crystal contacts, but also competing interactions that can hinder crystal assembly. The design of proteins that easily crystallize could also be used to validate and further enrich the microscopic insights obtained from the direct studies of protein-protein interactions~\cite{Lanci2012}.

Second, richer varieties of patchy models ought to be developed in order to address basic qualitative questions about protein crystal assembly. 
\begin{enumerate}
 \item What is the role of competing patches and dimer formation? Some proteins are observed to crystallize in more than one crystal lattice in the same solution composition~\cite{king:1956,king:1962,dixon:1992,faber:1990,mcree:1990,elgersma:1992}. The type of crystal that assembles can then depend on the initial protein concentration as well as on the experimental time and temperature.
\item What is the role of internal flexibility? The paradigm of a single well-folded protein structure is known to be overly simplistic~\cite{wright:2015}, but it remains a key requirement for protein crystal assembly. Internal flexibility of the protein may interfere with crystal assembly, but the physical constraints have not been given much attention.
\item What causes inverted solubility? Some proteins are characterized by a decreasing protein solubility with increasing temperature, i.e., an inverted solubility~\cite{veesler:2002,veesler:2003,veesler:2004}. Sometimes a single mutation can flip the solubility curve~\cite{mcmanus:2007,mcmanus:2015} or significantly change the assembly kinetics~\cite{asherie:2001,chen:2004}.  This phenomenon is tentatively attributed to the temperature dependence of the water entropy~\cite{gunton:2005,wentzel:2008}, but remains poorly understood overall. 
\item What are minimal models for membrane protein crystal assembly? As mentioned in the introduction, the crystallization of membrane proteins typically involves assembly principles that are different than for globular proteins~\cite{carpenter:2008,caffrey:2009}. Soft matter insights might be helpful in building better experimental guidance for this difficult, yet crucial~\cite{bhattacharya:2009}, problem. 
\end{enumerate}
The coming years will thus likely see the emergence of increasingly rich patchy models that can provide a clearer physical understanding of these and related processes.

In closing, it is important to note that patchy models have also found a use in the study of proteins beyond crystal assembly. The study of virus capsid formation, in particular, has greatly benefited~\cite{Hagan:2014}. The aggregation of proteins into amorphous structures, such as amyloids, is also within conceptual reach of similar approaches~\cite{bieler:2012,saric:2014}. The absence of clear microscopic information on protein-protein contacts in these systems, however, provides an additional challenge. The study of protein crystal assembly, which has the inherent advantage of providing structural feedback, when successful, may thus lead the way towards a better understanding of a broad array of protein-protein interactions.

\section*{Acknowledgments}
We acknowledge the help of our colleagues and collaborators at Duke and beyond, who have patiently walked us through various aspects of protein science over the years. We also gratefully acknowledge support from National Science Foundation Grant no. NSF DMR-1055586.

\appendix
\section{Characterizing crystal contacts: specific vs. non-specific protein-protein interactions}
Protein-protein interactions have often classified as being either \textit{specific} and \textit{non-specific}. Biological specificity is well-known to be a problematic label~\cite[Ch.~4]{Kay:2000}, but it is nonetheless a quite prevalent characterization, including of protein interactions that control crystal assembly. Its precise meaning, however, differs depending on the disciplinary context~(Table~\ref{table:sp_vs_non}). In this appendix, we aim to identify these different significations and thus provide a brief guide to the reader of the scientific literature on the topic.

In chemistry, specificity distinguishes certain attraction forces from others, although the classification of the various physical mechanisms is not unambiguous~\cite[(\S~18.8)]{israelachvili:1991}. In biophysics, the distinction between specific and non-specific interactions typically relies on the existence of an energy gap that clearly divides a single, strong (specific) interaction from the other (non-specific) ones~\cite{janin:1995,johnson:2011}. In molecular biology, specific interactions are deemed responsible for the stoichiometric recognition of a given target, while non-specific interactions arise from the promiscuous yet non-biologically relevant association of molecules~\cite{janin:1995b,carugo:1997,wilkinson:2004,zhuang:2011}. Specific interactions have thus been evolutionarily tuned to be free-energetically strong and geometrically oriented, while non-specific attractions have not. As mentioned above, this general weakness, however, may itself have evolved so as to prevent pathological aggregation~\cite{janin:1995,doye:2004}. Although these three definitions are not necessarily orthogonal to one another, we here specifically aim to clarify the last one. 
 
 \begin{table*}
  \caption{Summary of the properties that differentiate specific from non-specific interactions in different fields of study}
   \label{table:sp_vs_non}
 \centering
 \begin{tabular}{ccc}
 \hline
 \bf{Field}&\bf{Specific interactions}&\bf{Non-specific interactions}\\
 \hline
\multirow{2}{*}{Chemistry}&Hydrogen bonds,&Hydrophobic, depletion,\\
&salt bridges&van der Waals, electrostatic\\
\hline
 \multirow{2}{*}{Biophysics}&Unique, strong,&Many and weak\\
 &energetically gapped&\\
 \hline
\multirow{2}{*}{Molecular biology}&Evolutionary tuned, strong,&Weak,\\
&geometrically constrained&randomly distributed\\
 \hline
 \end{tabular}

 \end{table*}

When applied to crystal contacts, specificity has been used to suggest that these biologically non-functional interactions are in many ways indistinguishable from interfaces obtained by randomly bringing two proteins together~\cite{janin:1995}.
 These interfaces do present characteristics that are typical of non-specific interactions: they are weaker than functional interaction (of the order of few kJ/mol) and they do not have any obvious biological purpose. However, they also have unique properties that distinguish them from randomly selected surface patches~\cite{Derewenda2009,price:2009}. Specific local protein properties are correlated with crystallization and protein surface regions carrying such properties are more likely to be responsible for the interactions that drive crystal formation. Hence, crystal contacts are triggered by the basic chemical interactions present in any molecular system,  although they are typically different from the biologically functional ones. The importance of weak protein interactions is not limited to crystal contacts, but is also recognized in regard to the formation of transient protein complexes that are sometimes necessary for correct protein function~\cite{acuner:2011}.

\bibliographystyle{prsty}
\bibliography{References}

\begin{thebibliography}{100}

\bibitem{Note1}
This phenomenon is sometimes playfully referred to as the revenge of structural
  biology.

\bibitem{blundell:2002}
T.~L. Blundell, H. Jhoti, and C. Abell, Nat. Rev. Drug. Discov. {\bf 1},  45
  (2002).

\bibitem{kuhn:2002}
P. Kuhn, K. Wilson, M.~G. Patch, and R.~C. Stevens, Curr. Opin. Chem. Biol.
  {\bf 6},  704  (2002).

\bibitem{blundell:2004}
T.~L. Blundell and S. Patel, Curr. Opin. Pharmacol. {\bf 4},  490  (2004).

\bibitem{tickle:2004}
I. Tickle, A. Sharff, M. Vinkovic, J. Yon, and H. Jhoti, Chem. Soc. Rev. {\bf
  33},  558  (2004).

\bibitem{kitchen:2004}
D.~B. Kitchen, H. Decornez, J.~R. Furr, and J. Bajorath, Nat. Rev. Drug.
  Discov. {\bf 3},  935  (2004).

\bibitem{congreve:2005}
M. Congreve, C.~W. Murray, and T.~L. Blundell, Drug. Discov. Today {\bf 10},
  895  (2005).

\bibitem{huebsch:2009}
N. Huebsch and D.~J. Mooney, Nature {\bf 462},  426  (2009).

\bibitem{mcpherson:1999}
A. McPherson, {\em Crystallization of Biological Macromolecules} (CSHL Press,
  Cold Spring Harbor, 1999).

\bibitem{chayen:2008}
N.~E. Chayen and E. Saridakis, Nat. Meth. {\bf 5},  147  (2008).

\bibitem{Chayen2009}
N.~E. Chayen,  in {\em Advances in Protein Chemistry and Structural Biology},
  edited by J. Andrzej (Academic Press, London, 2009), Vol.~77, pp.\ 1--22.

\bibitem{Morange:2006}
M. Morange,  in {\em History and Epistemology of Molecular Biology and Beyond},
  edited by H. Rheinberger and S. Chadarevian (Max Planck Institute for the
  History of Science, ADDRESS, 2006), Vol.~310, pp.\ 179--188.

\bibitem{Khafizov:2014}
K. Khafizov, C. Madrid-Aliste, S.~C. Almo, and A. Fiser, Proc. Natl. Acad. Sci.
  U.S.A. {\bf 111},  3733  (2014).

\bibitem{jones:2014}
J. Nicola, Nature {\bf 505},  603  (2014).

\bibitem{chen:2004}
Q. Chen, P.~G. Vekilov, R.~L. Nagel, and R.~E. Hirsch, Biophys. J. {\bf 86},
  1702  (2004).

\bibitem{terwilliger:2009}
T.~C. Terwilliger, D. Stuart, and S. Yokoyama, Annu. Rev. Biophys. {\bf 38},
  371  (2009).

\bibitem{Note2}
The assembly of membrane proteins typically relies on different physical
  processes~\cite {caffrey:2009}. These processes are generally beyond the
  scope of this review, although we get back to this issue in the conclusion.

\bibitem{Dill:2012}
K.~A. Dill and J.~L. MacCallum, Science {\bf 338},  1042  (2012).

\bibitem{Grzesiek2009}
S. Grzesiek and H.-J. Sass, Curr. Opin. Struct. Biol. {\bf 19},  585  (2009).

\bibitem{kerfah:2015}
R. Kerfah, M.~J. Plevin, R. Sounier, P. Gans, and J. Boisbouvier, Curr. Opin.
  Struct. Biol. {\bf 32},  113   (2015).

\bibitem{doerr:2014}
A. Doerr, Nat. Meth. {\bf 11},  30  (2014).

\bibitem{li:2013}
X. Li, P. Mooney, S. Zheng, C.~R. Booth, M.~B. Braunfeld, S. Gubbens, D.~A.
  Agard, and Y. Cheng, Nat. Meth. {\bf 10},  584  (2013).

\bibitem{bai:2013}
X.-C. Bai, I.~S. Fernandez, G. McMullan, and S.~H. Scheres, eLife {\bf 2},
  e00461  (2013).

\bibitem{Yonekura:2015}
K. Yonekura, K. Kato, M. Ogasawara, M. Tomita, and C. Toyoshima, Proc. Nat.
  Acad. Sci. U.S.A. {\bf 112},  3368  (2015).

\bibitem{barends:2014}
T.~R. Barends {\it et~al.}, Nature {\bf 505},  244  (2014).

\bibitem{stevenson:2014}
H.~P. Stevenson {\it et~al.}, Proc. Natl. Acad. Sci. U.S.A. {\bf 111},  8470
  (2014).

\bibitem{blakeley:2008}
M.~P. Blakeley, P. Langan, N. Niimura, and A. Podjarny, Curr. Opin. Struct.
  Biol. {\bf 18},  593  (2008).

\bibitem{janin:1995}
J. Janin, Prog. Biophys. Mol. Biol. {\bf 64},  145  (1995).

\bibitem{doye:2004}
J.~P.~K. Doye, A.~A. Louis, and M. Vendruscolo, Phys. Biol. {\bf 1},  P9
  (2004).

\bibitem{doye:2006}
J.~P.~K. Doye and W.~C.~K. Poon, Curr. Opin. Colloid Interface Sci. {\bf 11},
  40  (2006).

\bibitem{ross:2004}
C.~A. Ross and M.~A. Poirier, Nat. Med. {\bf 10},  S10  (2004).

\bibitem{kmoch:2000}
S. Kmoch, J. Brynda, B. Asfaw, K. Bezouska, P. Novak, P. Rezacova, L. Ondrova,
  M. Filipec, J. Sedlacek, and M. Elleder, Hum. Mol. Gen. {\bf 9},  1779
  (2000).

\bibitem{pande:2001}
A. Pande, J. Pande, N. Asherie, A. Lomakin, O. Ogun, J. King, and G.~B.
  Benedek, Proc. Natl. Acad. Sci. U.S.A. {\bf 98},  6116  (2001).

\bibitem{evans:2004}
P. Evans, K. Wyatt, G.~J. Wistow, O.~A. Bateman, B.~A. Wallace, and C.
  Slingsby, J. Mol. Biol. {\bf 343},  435  (2004).

\bibitem{li:2008}
F.~F. Li, S.~Z. Wang, C. Gao, S.~G. Liu, B.~J. Zhao, M. Zhang, S.~Z. Huang,
  S.~Q. Zhu, and X. Ma, Mol. Vis. {\bf 14},  378  (2008).

\bibitem{newman:2012}
J. Newman {\it et~al.}, Acta Cryst. F {\bf 68},  253  (2012).

\bibitem{Bruno:2014}
A.~E. Bruno, A.~M. Ruby, J.~R. Luft, T.~D. Grant, J. Seetharaman, G.~T.
  Montelione, J.~F. Hunt, and E.~H. Snell, PLoS ONE {\bf 9},  e100782  (2014).

\bibitem{caffrey:2003}
M. Caffrey, J. Struct. Biol. {\bf 142},  108  (2003).

\bibitem{Stevens:2013}
H. Stevens, {\em Life Out of Sequence: A Data-Driven History of Bioinformatics}
  (University of Chicago Press, Chicago, 2013), p.\ 272.

\bibitem{derewenda:2010}
Z. Derewenda, Acta Cryst. D {\bf 66},  604  (2010).

\bibitem{anderson:2006}
M.~J. Anderson, C.~L. Hansen, and S.~R. Quake, Proc. Natl. Acad. Sci. U.S.A.
  {\bf 103},  16746  (2006).

\bibitem{saridakis:2009}
E. Saridakis and N.~E. Chayen, Trends Biotechnol. {\bf 27},  99  (2009).

\bibitem{dumetz:2009}
A.~C. Dumetz, A.~M. Chockla, E.~W. Kaler, and A.~M. Lenhoff, Cryst. Growth \&
  Des. {\bf 9},  682  (2009).

\bibitem{israelachvili:1991}
J.~N. Israelachvili, {\em Intermolecular and surface forces} (Academic Press,
  San Diego, 1991).

\bibitem{Chandler2005}
D. Chandler, Nature {\bf 437},  640  (2005).

\bibitem{gunton:2007}
J.~D. Gunton, A. Shiryayev, and D.~L. Pagan, {\em Protein Condensation}
  (Cambridge University Press, New York, 2007).

\bibitem{granick:2008}
S. Granick and S.~C. Bae, Science {\bf 322},  1477  (2008).

\bibitem{Berne2009}
B.~J. Berne, J.~D. Weeks, and R. Zhou, Annu. Rev. Phys. Chem. {\bf 60},  85
  (2009).

\bibitem{vekilov:2002}
P.~G. Vekilov and A.~A. Chernov, Solid State Phys. {\bf 57},  1  (2002).

\bibitem{devedjiev:2015}
Y.~D. Devedjiev, Acta Cryst. F {\bf 71},  157  (2015).

\bibitem{Matthews:1968}
B.~W. Matthews, J. Mol. Biol. {\bf 33},  491  (1968).

\bibitem{Malfois:1996}
M. Malfois, F. Bonnete, L. Belloni, and A. Tardieu, J. Chem. Phys. {\bf 105},
  3290  (1996).

\bibitem{lomakin:1996}
A. Lomakin, N. Asherie, and G.~B. Benedek, J. Chem. Phys. {\bf 104},  1646
  (1996).

\bibitem{muschol:1997}
M. Muschol and F. Rosenberger, J. Chem. Phys. {\bf 107},  1953  (1997).

\bibitem{stradner:2005}
A. Stradner, G.~M. Thurston, and P. Schurtenberger, J. Phys.: Condens. Matter
  {\bf 17},  S2805  (2005).

\bibitem{mcmanus:2007}
J.~J. McManus, A. Lomakin, O. Ogun, A. Pande, M. Basan, J. Pande, and G.~B.
  Benedek, Proc. Natl. Acad. Sci. U.S.A. {\bf 104},  16856  (2007).

\bibitem{talreja:2010}
S. Talreja, S.~L. Perry, S. Guha, V. Bhamidi, C.~F. Zukoski, and P.~J.~A.
  Kenis, J. Phys. Chem. B {\bf 114},  4432  (2010).

\bibitem{Gogelein2012}
C. G{\"o}gelein, D. Wagner, F. Cardinaux, G. N{\"a}gele, and S.~U. Egelhaaf, J.
  Chem. Phys. {\bf 136},  015102  (2012).

\bibitem{hansen:2006}
J.-P. Hansen and I.~R. McDonald, {\em Theory of simple liquids} (Academic
  Press, London, 2006).

\bibitem{asherie:2004}
N. Asherie, Methods {\bf 34},  266  (2004).

\bibitem{asherie:2012}
N. Asherie, Protein Pept. Lett. {\bf 19},  708  (2012).

\bibitem{sleutel:2015}
M. Sleutel, J.~F. Lutsko, D. Maes, and A.~E.~S. Van~Driessche, Phys. Rev. Lett.
  {\bf 114},  245501  (2015).

\bibitem{thomson:1987}
J.~A. Thomson, P. Schurtenberger, G.~M. Thurston, and G.~B. Benedek, Proc.
  Natl. Acad. Sci. U.S.A. {\bf 84},  7079  (1987).

\bibitem{dumetz:2008}
A.~C. Dumetz, A.~M. Chockla, E.~W. Kaler, and A.~M. Lenhoff, Biophys. J. {\bf
  94},  570  (2008).

\bibitem{Pusey:1986}
P.~N. Pusey and W. Vanmegen, Nature {\bf 320},  340  (1986).

\bibitem{Asakura:1954}
S. Asakura and F. Oosawa, J. Chem. Phys. {\bf 22},  1255  (1954).

\bibitem{Vrij:1976}
A. Vrij, Pure \& Appl. Chem. {\bf 48},  471  (1976).

\bibitem{gast:1983}
A.~P. Gast, C.~K. Hall, and W.~B. Russel, J. Colloid Interface Sci. {\bf 96},
  251  (1983).

\bibitem{Lekkerkerker:1992}
H.~N.~W. Lekkerkerker, W.~C.~K. Poon, P.~N. Pusey, A. Stroobants, and P.~B.
  Warren, Europhys. Lett. {\bf 20},  559  (1992).

\bibitem{Hagen1994}
M.~H.~J. Hagen and D. Frenkel, J. Chem. Phys. {\bf 101},  4093  (1994).

\bibitem{Asherie:1996}
N. Asherie, A. Lomakin, and G.~B. Benedek, Phys. Rev. Lett. {\bf 77},  4832
  (1996).

\bibitem{Tavares:1997}
F.~W. Tavares and S.~I. Sandler, {AIC}h{E} J. {\bf 43},  218  (1997).

\bibitem{Miller:2003}
M.~A. Miller and D. Frenkel, Phys. Rev. Lett. {\bf 90},  135702  (2003).

\bibitem{Liu:2005}
H. Liu, S. Garde, and S. Kumar, J. Chem. Phys. {\bf 123},  174505  (2005).

\bibitem{Pagan:2005}
D.~L. Pagan and J.~D. Gunton, J. Chem. Phys. {\bf 122},  184515  (2005).

\bibitem{lopezrendon:2006}
R. L\'opez-Rend\'on, Y. Reyes, and P. Orea, J. Chem. Phys. {\bf 125},  084508
  (2006).

\bibitem{Largo:2008}
J. Largo, M.~A. Miller, and F. Sciortino, J. Chem. Phys. {\bf 128},  134513
  (2008).

\bibitem{fortini:2008}
A. Fortini, E. Sanz, and M. Dijkstra, Phys. Rev. E {\bf 78},  041402  (2008).

\bibitem{Noro:2000}
M.~G. Noro and D. Frenkel, J. Chem. Phys. {\bf 113},  2941  (2000).

\bibitem{Foffi:2006}
G. Foffi and F. Sciortino, Phys. Rev. E {\bf 74},  050401(R)  (2006).

\bibitem{platten:2015}
F. Platten, N.~E. Valadez-P\'erez, R. Casta{\~n}eda-Priego, and S.~U. Egelhaaf,
  J. Chem. Phys. {\bf 142},  174905  (2015).

\bibitem{Wolde1997}
P.~R. ten Wolde and D. Frenkel, Science {\bf 277},  1975  (1997).

\bibitem{Foffi:2005}
G. Foffi, C. De~Michele, F. Sciortino, and P. Tartaglia, J. Chem. Phys. {\bf
  122},  224903  (2005).

\bibitem{Manley:2005}
S. Manley, H.~M. Wyss, K. Miyazaki, J.~C. Conrad, V. Trappe, L.~J. Kaufman,
  D.~R. Reichman, and D.~A. Weitz, Phys. Rev. Lett. {\bf 95},  238302  (2005).

\bibitem{Sastry:2006}
S. Sastry, Nature {\bf 441},  671  (2006).

\bibitem{Charbonneau:2007}
P. Charbonneau and D.~R. Reichman, Phys. Rev. E {\bf 75},  050401(R)  (2007).

\bibitem{Buzzaccaro:2007}
S. Buzzaccaro, R. Rusconi, and R. Piazza, Phys. Rev. Lett. {\bf 99},  098301
  (2007).

\bibitem{Lu2008}
P.~J. Lu, E. Zaccarelli, F. Ciulla, A.~B. Schofield, F. Sciortino, and D.~A.
  Weitz, Nature {\bf 453},  499  (2008).

\bibitem{rosenbaum:1996}
D. Rosenbaum, P.~C. Zamora, and C.~F. Zukoski, Phys. Rev. Lett. {\bf 76},  150
  (1996).

\bibitem{chayen:2005}
N.~E. Chayen, Progr. Biophys. Mol. Biol. {\bf 88},  329  (2005).

\bibitem{george:1994}
A. George and W.~W. Wilson, Acta Cryst. D {\bf 50},  361  (1994).

\bibitem{haas:1998}
C. Haas and J. Drenth, J. Phys. Chem. B {\bf 102},  4226  (1998).

\bibitem{wilson:2014}
W.~W. Wilson and L.~J. DeLucas, Acta Cryst. F {\bf 70},  543  (2014).

\bibitem{janin:1995b}
J. Janin and F. Rodier, Proteins {\bf 23},  580  (1995).

\bibitem{carugo:1997}
O. Carugo and P. Argos, Protein Sci. {\bf 6},  2261  (1997).

\bibitem{Pellicane:2004}
G. Pellicane, D. Costa, and C. Caccamo, J. Phys. Chem. B {\bf 108},  7538
  (2004).

\bibitem{stradner:2007}
A. Stradner, G. Foffi, N. Dorsaz, G. Thurston, and P. Schurtenberger, Phys.
  Rev. Lett. {\bf 99},  198103  (2007).

\bibitem{dorsaz:2009}
N. Dorsaz, G.~M. Thurston, A. Stradner, P. Schurtenberger, and G. Foffi, J.
  Phys. Chem. B {\bf 113},  1693  (2009).

\bibitem{foffi:2014}
G. Foffi, G. Savin, S. Bucciarelli, N. Dorsaz, G.~M. Thurston, A. Stradner, and
  P. Schurtenberger, Proc. Natl. Acad. Sci. U.S.A. {\bf 111},  16748  (2014).

\bibitem{chayen:2006}
N.~E. Chayen, E. Saridakis, and R.~P. Sear, Proc. Natl. Acad. Sci. U.S.A. {\bf
  103},  597  (2006).

\bibitem{Page:2009}
A.~J. Page and R.~P. Sear, J. Am. Chem. Soc. {\bf 131},  17550  (2009).

\bibitem{haas:1999}
C. Haas, J. Drenth, and W.~W. Wilson, J. Phys. Chem. B {\bf 103},  2808
  (1999).

\bibitem{curtis:2001}
R.~A. Curtis, H.~W. Blanch, and J.~M. Prausnitz, J. Phys. Chem. B {\bf 105},
  2445  (2001).

\bibitem{lomakin:1999}
A. Lomakin, N. Asherie, and G.~B. Benedek, Proc. Natl. Acad. Sci. U.S.A. {\bf
  96},  9465  (1999).

\bibitem{richards:1974}
F.~M. Richards, J. Mol. Biol. {\bf 82},  1   (1974).

\bibitem{liu:2007}
H. Liu, S.~K. Kumara, and F. Sciortino, J. Chem. Phys. {\bf 127},  084902
  (2007).

\bibitem{Bianchi2011}
E. Bianchi, R. Blaak, and C.~N. Likos, Phys. Chem. Chem. Phys. {\bf 13},  6397
  (2011).

\bibitem{mcmanus:2015}
S. James, M.~K. Quinn, and J.~J. McManus, Phys. Chem. Chem. Phys. {\bf 17},
  5413  (2015).

\bibitem{Derewenda2009}
M. Cie\'slik and Z.~S. Derewenda, Acta Cryst. D {\bf 65},  500  (2009).

\bibitem{price:2009}
W.~N. Price {\it et~al.}, Nat. Biotechnol. {\bf 27},  51  (2009).

\bibitem{Sear1999}
R.~P. Sear, J. Chem. Phys. {\bf 111},  4800  (1999).

\bibitem{chang:2004}
J. Chang, A.~M. Lenhoff, and S.~I. Sandler, J. Chem. Phys. {\bf 120},  3003
  (2004).

\bibitem{Gogelein2008}
C. G\"ogelein, G. N\"agele, R. Tuinier, T. Gibaud, A. Stradner, and P.
  Schurtenberger, J. Chem. Phys. {\bf 129},  085102  (2008).

\bibitem{neal:1998}
B.~L. Neal, D. Asthagiri, and A.~M. Lenhoff, Biophys. J. {\bf 75},  2469
  (1998).

\bibitem{hongjun:2009}
H. Liu, S.~K. Kumar, and J.~F. Douglas, Phys. Rev. Lett. {\bf 103},  018101
  (2009).

\bibitem{Whitelam2010}
S. Whitelam, Phys. Rev. Lett. {\bf 105},  088102  (2010).

\bibitem{Haxton2012}
T.~K. Haxton and S. Whitelam, Soft Matter {\bf 8},  3558  (2012).

\bibitem{bianchi:2006}
E. Bianchi, J. Largo, P. Tartaglia, E. Zaccarelli, and F. Sciortino, Phys. Rev.
  Lett. {\bf 97},  168301  (2006).

\bibitem{foffi:2007}
G. Foffi and F. Sciortino, J. Phys. Chem. B {\bf 111},  9702  (2007).

\bibitem{wertheim:1984a}
M.~S. Wertheim, J. Stat. Phys. {\bf 35},  19  (1984).

\bibitem{wertheim:1984b}
M.~S. Wertheim, J. Stat. Phys. {\bf 35},  35  (1984).

\bibitem{Bianchi:2007}
E. Bianchi, P. Tartaglia, E. La~Nave, and F. Sciortino, J. Phys. Chem. B {\bf
  111},  11765  (2007).

\bibitem{Bianchi:2008}
E. Bianchi, P. Tartaglia, E. Zaccarelli, and F. Sciortino, J. Chem. Phys. {\bf
  128},  144504  (2008).

\bibitem{Bianchi:2011}
E. Bianchi, R. Blaak, and C.~N. Likos, Phys. Chem. Chem. Phys. {\bf 13},  6397
  (2011).

\bibitem{fusco:2013b}
D. Fusco, J.~J. Headd, A. De~Simone, J. Wang, and P. Charbonneau, Soft Matter
  {\bf 10},  290  (2014).

\bibitem{glotzer:2007}
S.~C. Glotzer and M.~J. Solomon, Nat. Mater. {\bf 6},  557  (2007).

\bibitem{wang:2012}
Y. Wang, Y. Wang, D.~R. Breed, V.~N. Manoharan, L. Feng, A.~D. Hollingsworth,
  M. Weck, and D.~J. Pine, Nature {\bf 491},  51  (2012).

\bibitem{charbonneau:2007b}
P. Charbonneau and D. Frenkel, J. Chem. Phys. {\bf 126},  196101  (2007).

\bibitem{Romano2010}
F. Romano, E. Sanz, and F. Sciortino, J. Chem. Phys. {\bf 132},  184501
  (2010).

\bibitem{romano:2011}
F. Romano, E. Sanz, and F. Sciortino, J. Chem. Phys. {\bf 134},  174502
  (2011).

\bibitem{fantoni:2007}
R. Fantoni, D. Gazzillo, A. Giacometti, M.~A. Miller, and G. Pastore, J. Chem.
  Phys. {\bf 127},  234507  (2007).

\bibitem{granick:2009}
S. Granick, S. Jiang, and Q. Chen, Physics Today {\bf 2009},  68  (2009).

\bibitem{munao:2013}
G. Muna\`o, Z. Preisler, T. Vissers, F. Smallenburg, and F. Sciortino, Soft
  Matter {\bf 9},  2652  (2013).

\bibitem{ruzicka:2010}
B. Ruzicka, E. Zaccarelli, L. Zulian, R. Angelini, M. Sztucki, A. Moussaid, T.
  Narayanan, and F. Sciortino, Nat. Mat. {\bf 10},  56  (2010).

\bibitem{krissinel:2010}
E. Krissinel, J. Comput. Chem. {\bf 31},  133  (2010).

\bibitem{wukovitz:1995}
S.~W. Wukovitz and T.~O. Yeates, Nat. Struct. Biol. {\bf 2},  1062  (1995).

\bibitem{dasgupta:1997}
S. Dasgupta, G.~H. Iyer, S.~H. Bryant, C.~E. Lawrence, and J.~A. Bell,
  Proteins: Structure, Function, and Bioinformatics {\bf 28},  494  (1997).

\bibitem{minor:2008}
M. Chruszcz, W. Potrzebowski, M.~D. Zimmerman, M. Grabowski, H. Zheng, P.
  Lasota, and W. Minor, Prot. Sci. {\bf 17},  623  (2008).

\bibitem{fusco:2014b}
D. Fusco and P. Charbonneau, J. Phys. Chem. B {\bf 118},  8034  (2014).

\bibitem{hloucha:2001}
M. Hloucha, J.~F.~M. Lodge, A.~M. Lenhoff, and S.~I. Sandler, J. Cryst. Growth
  {\bf 232},  195  (2001).

\bibitem{dixit:2002}
N.~M. Dixit and C.~F. Zukoski, J. Chem. Phys. {\bf 117},  8540  (2002).

\bibitem{dorsaz:2012}
N. Dorsaz, L. Filion, F. Smallenburg, and D. Frenkel, Farad. Disc. {\bf 159},
  9  (2012).

\bibitem{fusco:2013a}
D. Fusco and P. Charbonneau, Phys. Rev. E {\bf 88},  012721  (2013).

\bibitem{Price:2011}
N.~W. Price~II {\it et~al.}, Microb. Inform. Exp. {\bf 1},  1  (2011).

\bibitem{Pellicane2008}
G. Pellicane, G. Smith, and L. Sarkisov, Phys. Rev. Lett. {\bf 101},  248102
  (2008).

\bibitem{Taudt:2015}
A. Taudt, A. Arnold, and J. Pleiss, Phys. Rev. E {\bf 91},  033311  (2015).

\bibitem{Schmit:2012}
J.~D. Schmit and K. Dill, J. Am. Chem. Soc. {\bf 134},  3934  (2012).

\bibitem{Liu:2012}
Z. Liu, W.-P. Zhang, Q. Xing, X. Ren, M. Liu, and C. Tang, Angew. Chem. Intl.
  Ed. {\bf 51},  469  (2012).

\bibitem{acuner:2011}
S.~E. Acuner~Ozbabacan, H.~B. Engin, A. Gursoy, and O. Keskin, Protein Eng.
  Des. Sel {\bf 24},  635  (2011).

\bibitem{Lanci2012}
C.~J. Lanci, C.~M. MacDermaid, S.-g. Kang, R. Acharya, B. North, X. Yang, X.~J.
  Qiu, W.~F. DeGrado, and J.~G. Saven, Proc. Natl. Acad. Sci. U.S.A. {\bf 109},
   7304  (2012).

\bibitem{king:1956}
M.~V. King, B.~S. Magdoff, M.~B. Adelman, and D. Harker, Acta Cryst. {\bf 9},
  460  (1956).

\bibitem{king:1962}
M.~V. King, J. Bello, E.~H. Pignataro, and D. Harker, Acta Cryst. {\bf 15},
  144  (1962).

\bibitem{dixon:1992}
M. Dixon, H. Nicholson, L. Shewchuk, W. Baase, and B. Matthews, J. Mol. Biol.
  {\bf 227},  917  (1992).

\bibitem{faber:1990}
H. Faber and B. Matthews, Nature {\bf 348},  263  (1990).

\bibitem{mcree:1990}
D.~E. McRee, S.~M. Redford, T.~E. Meyer, and M.~A. Cusanovich, J. Biol. Chem.
  {\bf 265},  5364  (1990).

\bibitem{elgersma:1992}
A.~V. Elgersma, M. Ataka, and T. Katsura, J. Cryst. Growth {\bf 122},  31
  (1992).

\bibitem{wright:2015}
P.~E. Wright and H.~J. Dyson, Nature Rev. Mol. Cell. Biol. {\bf 16},  18
  (2015).

\bibitem{veesler:2002}
M. Budayova-Spano, F. Bonneté, J.-P. Astier, and S. Veesler, J. Cryst. Growth
  {\bf 235},  547  (2002).

\bibitem{veesler:2003}
M. Budayova-Spano, S. Lafont, J.-P. Astier, C. Ebel, and S. Veesler, J. Cryst.
  Growth {\bf 217},  311  (2000).

\bibitem{veesler:2004}
S. Veesler, N. Fertè, M.-S. Costes, M. Czjzek, and J.-P. Astier, Cryst. Growth
  \& Design {\bf 4},  1137  (2004).

\bibitem{asherie:2001}
N. Asherie, J. Pande, A. Pande, J.~A. Zarutskie, J. Lomakin, A. Lomakin, O.
  Ogun, L.~J. Stern, J. King, and G.~B. Benedek, J. Mol. Biol. {\bf 314},  663
  (2001).

\bibitem{gunton:2005}
A. Shiryayev, D.~L. Pagan, J.~D. Gunton, D.~S. Rhen, A. Saxena, and T. Lookman,
  J. Chem. Phys. {\bf 122},  234911  (2005).

\bibitem{wentzel:2008}
N. Wentzel and J.~D. Gunton, J. Phys. Chem. B {\bf 112},  7803  (2008).

\bibitem{carpenter:2008}
E.~P. Carpenter, K. Beis, A.~D. Cameron, and S. Iwata, Curr. Opin. Struct.
  Biol. {\bf 18},  581  (2008).

\bibitem{caffrey:2009}
M. Caffrey, Annu. Rev. Biophys. {\bf 38},  29  (2009).

\bibitem{bhattacharya:2009}
A. Bhattacharya, Nature {\bf 459},  24  (2009).

\bibitem{Hagan:2014}
M.~F. Hagan, Adv. Chem. Phys. {\bf 155},  1  (2014).

\bibitem{bieler:2012}
N.~S. Bieler, T.~P.~J. Knowles, D. Frenkel, and R. V\'acha, PLoS Comput. Biol.
  {\bf 8},  e1002692  (2012).

\bibitem{saric:2014}
A. {\v{S}}ari\'c, Y.~C. Chebaro, T.~P.~J. Knowles, and D. Frenkel, Proc. Natl.
  Acad. Sci. U.S.A. {\bf 111},  17869  (2014).

\bibitem{Kay:2000}
L.~E. Kay, {\em Who Wrote the Book of Life?: A History of the Genetic Code}
  (Stanford University Press, Stanford, 2000).

\bibitem{johnson:2011}
M.~E. Johnson and G. Hummer, Proc. Natl. Acad. Sci. U.S.A. {\bf 108},  603
  (2011).

\bibitem{wilkinson:2004}
K.~D. Wilkinson,  in {\em Protein-Protein Interactions: Methods and Protocols}
  (Humana Press, New York, 2004), Vol.~261, pp.\ 15--31.

\bibitem{zhuang:2011}
T. Zhuang, B.~K. Jap, and C.~R. Sanders, J. Am. Chem. Soc. {\bf 133},  20571
  (2011).

\end{thebibliography}

\end{document}